\documentclass{article}
\usepackage{authblk}
\usepackage{graphicx}
\usepackage{float}
\usepackage{color}

\RequirePackage{caption}
\RequirePackage{cite}

\RequirePackage{jabbrv}
\RequirePackage{newtxtext,newtxmath,courier}

\title{Group-velocity symmetry in photonic crystal fibre for ultra-tunable quantum frequency conversion}

\author[1]{C.~Parry$^{\ast}$}
\author[1]{P.~B.~Main,}
\author[1]{T.~A.~Wright}
\author[1]{P.~J.~Mosley}

\affil[1]{Centre for Photonics and Photonic Materials, Department of Physics, University of Bath, Bath, BA2 7AY, UK,
{$^{\ast}$c.parry@bath.ac.uk}
}

\begin{document}
\date{}
\maketitle
\begin{abstract}
Low-noise frequency conversion of single photons is a critical tool in establishing fibre-based quantum networks. We show that a single photonic crystal fibre can achieve frequency conversion by Bragg-scattering four-wave mixing of source photons from an ultra-broad wavelength range by engineering a symmetric group velocity profile. Furthermore, we discuss how pump tuning can mitigate realistic discrepancies in device fabrication. This enables a single highly adaptable frequency conversion interface to link disparate nodes in a quantum network via the telecoms band.
\end{abstract}

\section{Introduction}

The ability to transmit quantum information and distribute entanglement between nodes is a prerequisite for modular networks capable of large-scale quantum information processing and quantum communication \cite{kimble2008quantum, walmsley2016building}. Single photons enable the robust transfer of quantum information due to their weak interaction with the environment. They are therefore well-established information carriers within a variety of platforms including networks of trapped-ions \cite{Stephenson2020High-Rate-High-Fidelity-Entanglement}, solid-state spin qubits \cite{Humphreys2018Deterministic-delivery-of-remote}, and quantum key distribution systems both via metropolitan fibre networks \cite{Bunandar2018Metropolitan-Quantum-Key-Distribution} and satellite relays \cite{Liao2017Satellite-to-ground-quantum-key-distribution}. Many techniques have matured to the point at which high-quality single photons can be produced across a wide range of wavelengths and bandwidths, but as a result not all photonic technologies are compatible. Therefore, low-noise frequency conversion of single photons is a critical tool in establishing universal quantum interconnects between information storage and processing nodes \cite{Awschalom2020Development-of-Quantum-InterConnects}. Not only can conversion to telecommunication bands enable efficient communication between nodes whose operating wavelengths are not compatible with low-loss transmission in fiber \cite{Walker2018Long-Distance-Single-Photon}, but also systems that emit at different wavelengths may be shifted into resonance with one another to form hybrid networks or to compensate for discrepancies in fabrication \cite{Rakher2010Quantum-transduction-of-telecommunications-band}. Additionally, more general transformations of the spectral-temporal mode such as bandwidth manipulation enable coupling between systems with widely varying emission and absorption profiles \cite{Lavoie2013,Karpinski2017}, as well as the coherent control of time bin-encoded quantum information \cite{Donohue2013,Donohue2014}. These applications share the need for a versatile frequency conversion apparatus with a broad spectral acceptance window.

Frequency conversion of single photons has previously been achieved through electro-optic methods \cite{Wright2017,Karpinski2017} and parametric interactions in nonlinear optical materials driven by high-intensity laser fields. Second-order nonlinearities are used to yield large frequency shifts by sum- and difference-frequency generation (SFG and DFG). In this way, frequency conversion to the telecoms C-band has been demonstrated with ionic emission lines \cite{Rutz2017Quantum-Frequency-Conversion, wright2018two}, atomic ensembles \cite{Albrecht2014A-waveguide-frequency-converter, Ikuta2018Polarization-insensitive-frequency}, quantum dots \cite{morrison2021bright}, and using photons emitted by diamond colour centres \cite{Tchebotareva2019Entanglement-between-a-Diamond}. The wavelengths from and to which photons are converted, referred to here as the source and target respectively, are fixed by the wavelength of the pump laser and phase matching in the nonlinear crystal. There can be the potential for a small amount of tuning in nonlinear crystals, although changes to the source and target wavelengths are typically small for a given device \cite{Maring2018Quantum-frequency-conversion,siverns2019neutral,Zaske2012Visible-to-Telecom-Quantum-Frequency,Fisher2020Integrated-Optical-Device}. Second-order nonlinearities in bulk crystals have also been used to demonstrate bandwidth manipulation (accompanied by a significant spectral shift) of heralded single photons \cite{Donohue2013,Donohue2014}.

On the other hand, processes mediated by third-order optical nonlinearities present more opportunities for spectral manipulation of single photons. Cross-phase modulation has been used to imprint a time lens to achieve bandwidth manipulation of single photons in optical fiber \cite{Hirooka2006,Matsuda2016}, and four-wave mixing processes provide a flexible means of achieving smaller frequency shifts where pump lasers for SFG or DFG are not readily available, for example from the near infrared to the telecoms C-band \cite{Bock2018High-fidelity-entanglement-between,aso2000broadband}. Specifically, Bragg-scattering four-wave mixing (BS-FWM) enables photons to be frequency converted by an amount equal to the difference between two pump fields \cite{mcguinness2010quantum} with the potential to engineer uni-directional conversion \cite{Bell2017Uni-directional-wavelength-conversion} and achieve both high conversion efficiency \cite{Lefrancois2015Optimizing-optical-Bragg} and low noise \cite{Li2016Efficient-and-low-noise-single-photon-level}. However, the prospect of more general spectral mode conversion by four-wave mixing, such as bandwidth manipulation, has hitherto been obstructed by narrow conversion bandwidths and group velocity walkoff, and hence is yet to be demonstrated experimentally.

Typically, a specific BS-FWM device, such as a waveguide or fibre, will enable two pump fields with wavelengths $\lambda_\text{P}$ and $\lambda_\text{Q}$ to shift a source photon from wavelength $\lambda_\text{S}$ to a target wavelength $\lambda_\text{T}$. High-efficiency conversion -- in theory up to 100\% -- can be achieved at wavelengths for which perfect phase matching is achieved \cite{Lefrancois2015Optimizing-optical-Bragg} and has been demonstrated for small frequency shifts within the telecoms C-band \cite{Clark2013High-efficiency-frequency-conversion}. However, any change to the operating wavelengths will rapidly degrade performance. For example, if the source wavelength were to change, phase mismatch would accrue and, even if one pump field were tuned to maintain energy conservation, conversion efficiency would drop rapidly and a new device with a different dispersion profile would be required. Hence, typical conversion bandwidths in conventional fibre are on the order of a few nanometers \cite{Anjum2019Bandwidth-enhancement-of-inter-modal}, though we note that work has been done to engineer larger conversion bandwidths in the 10--100\,nm range both in BS-FWM conversion in the general case of waveguided modes \cite{Christensen2018Shape-preserving-and-unidirectional-frequency} and in silicon nitride microresonators \cite{Singh2019Quantum-frequency-conversion} as well as Raman-mediated frequency conversion in atomic ensembles \cite{Bustard2017Quantum-frequency-conversion}.

Photonic crystal fibre (PCF), in which a matrix of air holes surrounds a silica glass core, is an ideal platform for BS-FWM due to the tight modal confinement and low loss which enable moderate laser powers to produce high intensities that are maintained over long interaction lengths. Although the structure of these fibres can be designed to alter their dispersion and tailor the phase matching of BS-FWM to a particular set of desired wavelengths, PCF typically suffers from the same limited conversion bandwidth as other platforms used for BS-FWM \cite{Wright2020Resource-efficient-frequency-conversion}.

In this paper we demonstrate that dispersion-engineered PCF enables BS-FWM of an unprecedentedly wide range of source photon wavelengths to a narrow target band. We show how the unique characteristics of PCF enable its group velocity to be modified such that efficient frequency conversion to 1550\,nm can be maintained across source wavelengths throughout the near infrared. We also show that the acceptance bandwidth, which is significantly wider than any other practical BS-FWM platform, is rendered robust against small discrepancies in fabrication through tuning of the pump wavelengths. This allows the construction of a single, highly versatile PCF device capable of unifying in the telecoms C-band input photons from a range of hundreds of nanometers, paving the way to a universal quantum frequency conversion interface to link disparate nodes in a quantum network or compensate for the dispersion of emission wavelength in single-photon emitters such as quantum dots \cite{senellart2017high}.

\section{Group velocity symmetry in BS-FWM}

The key requirement for the frequency conversion of a broad range of source wavelengths -- group velocity that is symmetric across a range of source and pump frequencies -- can be derived from the energy conservation and phase matching conditions needed for efficient BS-FWM.

\begin{figure}
\centering
\includegraphics[width = \textwidth]{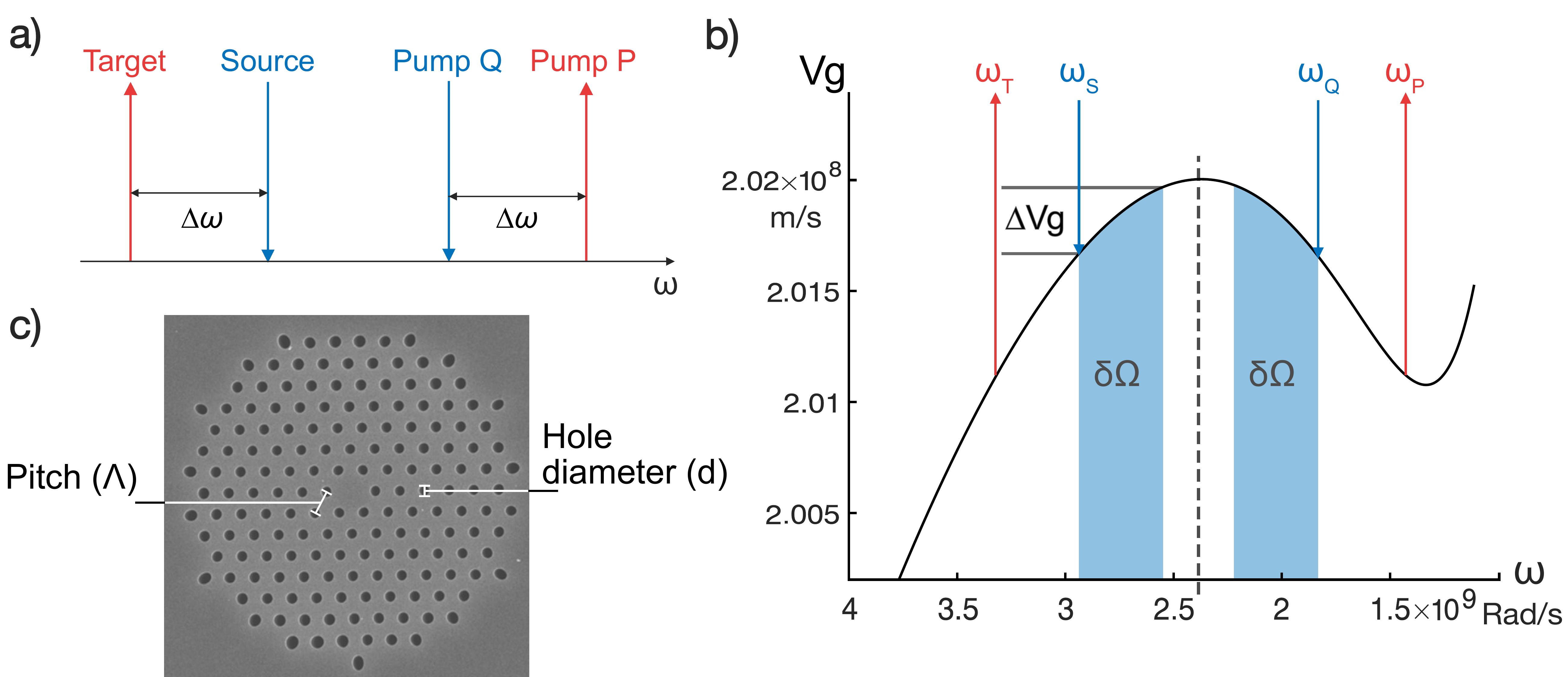}
\caption{\footnotesize (a) Bragg-scattering four-wave mixing in which source photons are converted to a target frequency through the annihilation and creation of photons in each of two pump fields P and Q respectively. The frequency difference between the pump fields is matched by that between the source and target. (b) Symmetric group velocity distribution, illustrated here for a PCF with a pitch of 1.39$\mu\mathrm{m}$, $\frac{\mathrm{d}}{\Lambda}$ of 0.55, and zero dispersion wavelength of 794$\mathrm{nm}$; frequencies are overlaid. A change $\delta\Omega$ in the source frequency, $\omega_{\,\mathrm{S}}$, requires a corresponding shift of $-\delta\Omega$ in the frequency of pump Q, $\omega_{\,\mathrm{Q}}$, and results in an equal shift, $\mathrm{\Delta v_g}$, in the group velocities  of both the source and long-wavelength pump Q.}
\label{fig:bs-fwm}
\end{figure}

Consider a set of frequencies $\{\omega_\text{p}, \omega_\text{q}, \omega_\text{s}, \omega_\text{t}\}$ at which a fixed-wavelength pump (p) and and tunable pump (q) convert a variable-wavelength source photon (s) to a fixed target frequency $\omega_\text{t}$. For efficient conversion, the interaction must satisfy energy conservation:
\begin{equation}
\omega_\text{p} + \omega_\text{t} = \omega_\text{q} + \omega_\text{s}, 
\label{eq:energy}
\end{equation}
and the phase mismatch, $\Delta\kappa_\text{BS}$, between the fields must be minimised \cite{McKinstrie2005Translation-of-quantum-states}:
\begin{equation}
\Delta\kappa_\text{BS} = \frac{1}{2}\Delta\beta + \frac{1}{2}\gamma\left(P_\text{q} - P_\text{p}\right).
\label{eq:mom}
\end{equation}
Here the linear phase mismatch is $\Delta\beta = \beta_\text{p} + \beta_\text{t} - \beta_\text{q} + \beta_\text{s}$ and the propagation constants, $\beta_j = \beta(\omega_j)$ with $j = \{\text{p}, \text{q}, \text{s}, \text{t}\}$, are defined as the longitudinal component of the wave vector in the fundamental mode of an optical fibre. The correction due to phase modulation introduced by nonlinearity $\gamma$ is typically small enough to be ignored at the frequency detunings we consider, particularly as the peak powers of the two pumps, $P_\text{p}$ and $P_\text{q}$, required for BS-FWM conversion are often similar; hence we target interactions with $\Delta\beta = 0$ for perfect phase matching.

For conversion to a fixed target frequency, $\omega_\text{t}$, a small change in the frequency of the source photon, $\delta\omega_\text{s}$, requires a shift in the detuning between the two pump frequencies to maintain energy conservation. We cannot change the frequency of the fixed-wavelength pump, so the required shift must be in the frequency of the tunable pump, altering it by $\delta\omega_\text{q}$. In order to maintain energy conservation:
\begin{equation}
\omega_\text{p} - (\omega_\text{q} + \delta\omega_\text{q}) - (\omega_\text{s} + \delta\omega_\text{s}) + \omega_\text{t} = 0,
\end{equation}
\noindent
and therefore by substitution from Eq.\,\ref{eq:energy}, $\delta\omega_\text{q} = - \delta\omega_\text{s} \equiv - \delta\Omega$. To find the corresponding condition to maintain phase matching we expand $\beta$ about $\omega_\text{s}$:
\begin{equation}
\beta(\omega_\text{s} + \delta\Omega) = \beta_\text{s} + \beta_\text{1}(\omega_\text{s}) \delta\Omega + \mathcal{O}(\delta\Omega^2), \hspace{5mm} \text{and} \hspace{3mm} \beta_1(\omega_\text{s}) = \left. \frac{\partial\beta}{\partial\omega}\right\vert_{\omega_\text{s}} = \frac{1}{v_g(\omega_\text{s})},
\end{equation}
where $v_g(\omega_\text{s})$ denotes the group velocity at frequency $\omega_\text{s}$. A similar expansion about $\omega_\text{q}$ yields the Bragg scattering phase matching condition:
\begin{equation}
\beta_{p} - (\beta_{q} - \beta_{1}(\omega_\text{q})\delta\Omega) - (\beta_{s} + \beta_{1}(\omega_\text{s})\delta \Omega) + \beta_{t} \approx 0,
\end{equation}
which becomes exact in the limit $\delta\Omega \rightarrow 0$. By comparison with Eq.\,\ref{eq:mom} we see that $\beta_{1}(\omega_\text{q}) = \beta_{1}(\omega_\text{s})$ to maintain phase matching, and that this condition is satisfied by matching the group velocity of pump Q with that of the source photon: $v_g(\omega_\text{q}) = v_g(\omega_\text{s})$. It follows that if group velocity matching can be sustained as $\delta\Omega$ increases, such that $v_g(\omega_\text{q} - \delta\Omega) = v_g(\omega_\text{s} + \delta\Omega)$, then phase matched BS-FWM frequency conversion across the whole range of source frequency $\delta\Omega$ becomes possible.

As shown in Fig.\,\ref{fig:bs-fwm}b, if we begin with $\omega_\text{s} = \omega_\text{q}$ at the zero-dispersion wavelength (ZDW), $\lambda_0 = 2 \pi c / \omega_0$, it becomes possible to increase $\omega_\text{q}$ by $\delta\Omega$ and decrease $\omega_\text{s}$ by $\delta\Omega$ while maintaining equal group velocity $v_g(\omega_\text{q} - \delta\Omega) = v_g(\omega_\text{s} + \delta\Omega)$ if, and only if, the group velocity profile of the fibre is symmetric about $\omega_0$. Hence imposing symmetry of the group velocity enables source photons across a broad range of wavelengths to be converted to to a single target frequency as long as the tunable pump frequency $\omega_\text{q}$ can be adjusted suitably to satisfy energy conservation. The dispersion in combination with the fixed pump frequency $\omega_\text{p}$ determines the target wavelength.

\section{Widely tunable BS-FWM in PCF}

The condition of group velocity symmetry is not generally satisfied in dispersive media, however, it can be achieved across a broad range of wavelengths using the flexible dispersion control of PCF. The propagation constant in PCF may be expanded about a zero-dispersion point $\omega_0$ as:
\begin{equation}
\beta (\omega)=\beta_0 + (\omega-\omega_0)\beta_1+ \frac{1}{2}(\omega-\omega_0)^2 \beta_2 + \frac{1}{6} (\omega-\omega_0)^3 \beta_3 + \frac{1}{24} (\omega-\omega_0)^4 \beta_4 \ldots  
\end{equation}
where
\begin{equation}
\beta_n = \left. \frac{\partial^n\beta}{\partial\omega^n}\right\vert_{\omega_0}.
\end{equation}
Considering a similar expansion of the inverse group velocity:
\begin{equation}
\frac{1}{v_g(\omega)} = \beta_1(\omega) = \beta_1 + (\omega-\omega_0)\beta_2 + \frac{1}{2} (\omega-\omega_0)^2 \beta_3 + \frac{1}{6}  (\omega-\omega_0)^3 \beta_4 \ldots 
\end{equation}
we see straightforwardly that odd-order terms in the expansion of the propagation constant, $\{\beta_1, \beta_3, \beta_5\ldots\}$, yield a symmetric group velocity profile, whereas even-order terms, $\{\beta_2, \beta_4\ldots\}$ are antisymmetric in group velocity. Therefore to ensure group velocity symmetry, and hence phase matched frequency conversion, it is necessary to minimise the contribution from even-order terms over as large a bandwidth as possible.

\begin{figure}[h]
\centering
\includegraphics[width = 0.8\textwidth]{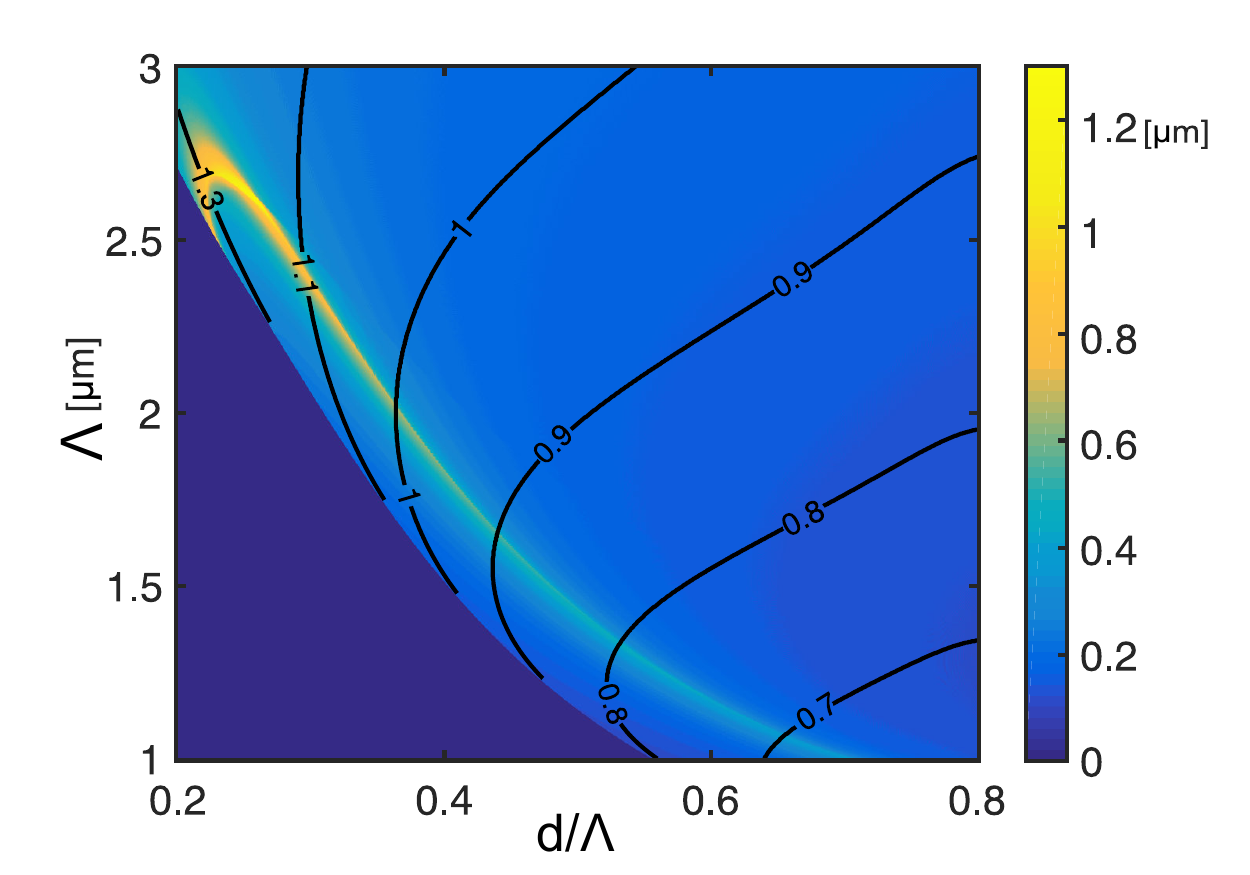}
\caption{Calculated group velocity symmetry about the first zero dispersion wavelength for a range of PCF design parameters of pitch, $\Lambda$, and ratio of hole diameter to pitch, $d/\Lambda$. Colours indicate the bandwidth in $\upmu$m over which the group velocity is symmetric in frequency such that $\Delta v_g = | v_g(\omega_\text{q} - \delta\Omega) - v_g(\omega_\text{s} + \delta\Omega) | < 5\times10^7 \text{ms}^{-1}$. Contours indicate the zero dispersion wavelength in $\upmu$m.}
\label{fig:symmetric}
\end{figure}

In order to assess the capability of PCF to achieve highly tunable BS-FWM frequency conversion, we make use of a set of empirical relationships that describe propagation in the fundamental mode set out by Saitoh and Koshiba \cite{saitoh2005empirical}. For regular PCF structures, these relationships parameterise the frequency dependence of the effective refractive index, $n_\text{eff}(\omega)$, and thence the propagation constant, $\beta(\omega) = n_\text{eff}(\omega)\omega/c$ in terms of only two variables: the distance between the air holes, known as the pitch ($\Lambda$), and their diameter ($d$) which is typically scaled to the pitch, $d/\Lambda$. Adjusting these two parameters enables the relative contributions from even and odd terms in the expansion of $\beta(\omega)$ to be controlled to a certain extent. 

Fig.\,\ref{fig:symmetric} shows the optimal combinations of $\Lambda$ and $d/\Lambda$ for group velocity symmetry over a wide range of PCF designs. The contours map how the ZDW changes over the parameter space, and the colour map indicates the level of group velocity symmetry. PCF designs that exhibit a useful level of group velocity symmetry are indicated by calculating the largest detuning around the ZDW at which the absolute value of the difference in group velocities, $\Delta v_g = | v_g(\omega_\text{q} - \delta\Omega) - v_g(\omega_\text{s} + \delta\Omega) |$, is less than $5\times10^7 \text{ms}^{-1}$ (this value was chosen simply as an illustration, but represents the group velocity shift one might expect from a 1\% change in structural parameters of a typical PCF used for BS-FWM). The blank area in the lower left quadrant is beyond the limit of validity of the empirical relationships. The data in Fig.\,\ref{fig:symmetric} enable us to select PCF designs that exhibit a high level of group velocity symmetry and hence suitability for ultra-tunable BS-FWM.

\begin{figure}[h!]
\centering
\includegraphics[width = 0.9\textwidth]{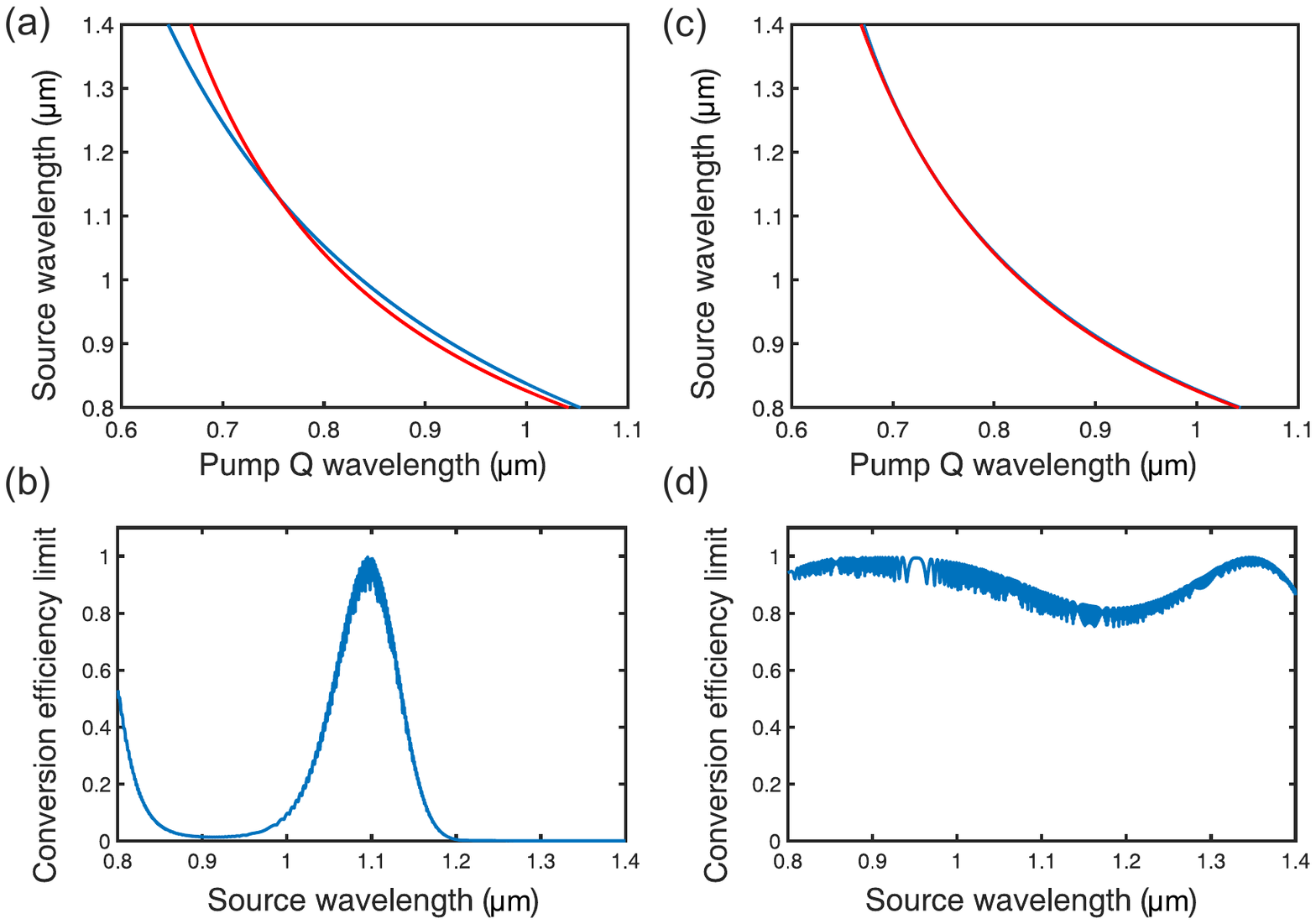}
\caption{BS-FWM in (a,b) a typical PCF without group velocity symmetry where $d = 2.35\,\upmu$m and $d/\Lambda = 0.5$ and (c,d) a PCF exhibiting high group velocity symmetry where $\Lambda = 1.78\,\upmu$m and $d/\Lambda = 0.437$. (a,c) show the locus of zero phase mismatch (blue) and energy conservation for conversion to $\lambda_\text{t} = 1550$\,nm (red). (b,d) show the maximum overlap between phase matching and pump functions as function of source wavelength, indicating maximum possible conversion efficiency to $\lambda_\text{t} = 1550$\,nm. The pump FWHM bandwidth in both cases is 5\,nm. Rapid oscillations in (b, d) are numerical artefacts resulting from limited wavelength resolution.}
\label{fig:sym_vs_non}
\end{figure}
In Fig.\,\ref{fig:sym_vs_non} we make a comparison between BS-FWM in a PCF that exhibits a high degree of group velocity symmetry around a ZDW of 900\,nm and a typical PCF without group velocity symmetry. We plot contours of the maximum of the BS-FWM phase matching intensity \cite{Garay-Palmett2007Photon-pair-state-preparation}:
\begin{equation}
\Phi = \left\vert \text{sinc}\left(\frac{\Delta\kappa_\text{BS} L}{2} \right) \right\vert^2
\end{equation}
with contours depicting perfect energy conservation for frequency conversion to a single target wavelength of 1550\,nm by a fixed pump $\lambda_\text{p}$ overlaid. We see that for the typical PCF without group velocity symmetry, the contours of zero phase mismatch and energy conservation intersect at a single pair of $\lambda_\text{q}$ and $\lambda_\text{s}$. However, the PCF with a high degree of group velocity symmetry matches the loci of points satisfying both conditions over a much larger wavelength range. Multiplying $\Phi$ by a Gaussian frequency distribution $\alpha_p$ with bandwidth $\sigma_\text{p}$ around central frequency $\omega_\text{p0}$ that describes the fixed-wavelength pump p:
\begin{equation}
\alpha_\text{p} = \exp{\left[\left(\frac{\omega_\text{p} - \omega_\text{p0}}{\sigma_\text{p}}\right)^2\right]} = \exp{\left[\left(\frac{\omega_\text{q} + \omega_\text{s} - \omega_\text{t} - \omega_\text{p0}}{\sigma_\text{p}}\right)^2\right]}
\end{equation}
 yields the maximum conversion efficiency that could be achieved in either configuration. Hence we see in Fig.\,\ref{fig:sym_vs_non} (b) for the typical PCF that the possible conversion efficiency is high for only a small range of source wavelengths. On the other hand, the PCF that exhibits a high degree of group velocity symmetry results in wide-ranging overlap between the phase matching and fixed-pump functions and high conversion efficiency over a broad wavelength range as shown in Fig.\ref{fig:sym_vs_non} (d). Therefore we see that, by maintaining group velocity symmetry over an unusually wide range of wavelengths, simply through tuning $\lambda_\text{q}$ any source wavelength between 800\,nm and 1.3\,$\upmu$m may be efficiently converted to the same target wavelength at 1.55\,$\upmu$m in a single section of PCF.

\begin{figure}[h!]
\centering
\includegraphics[width = 0.85\textwidth]{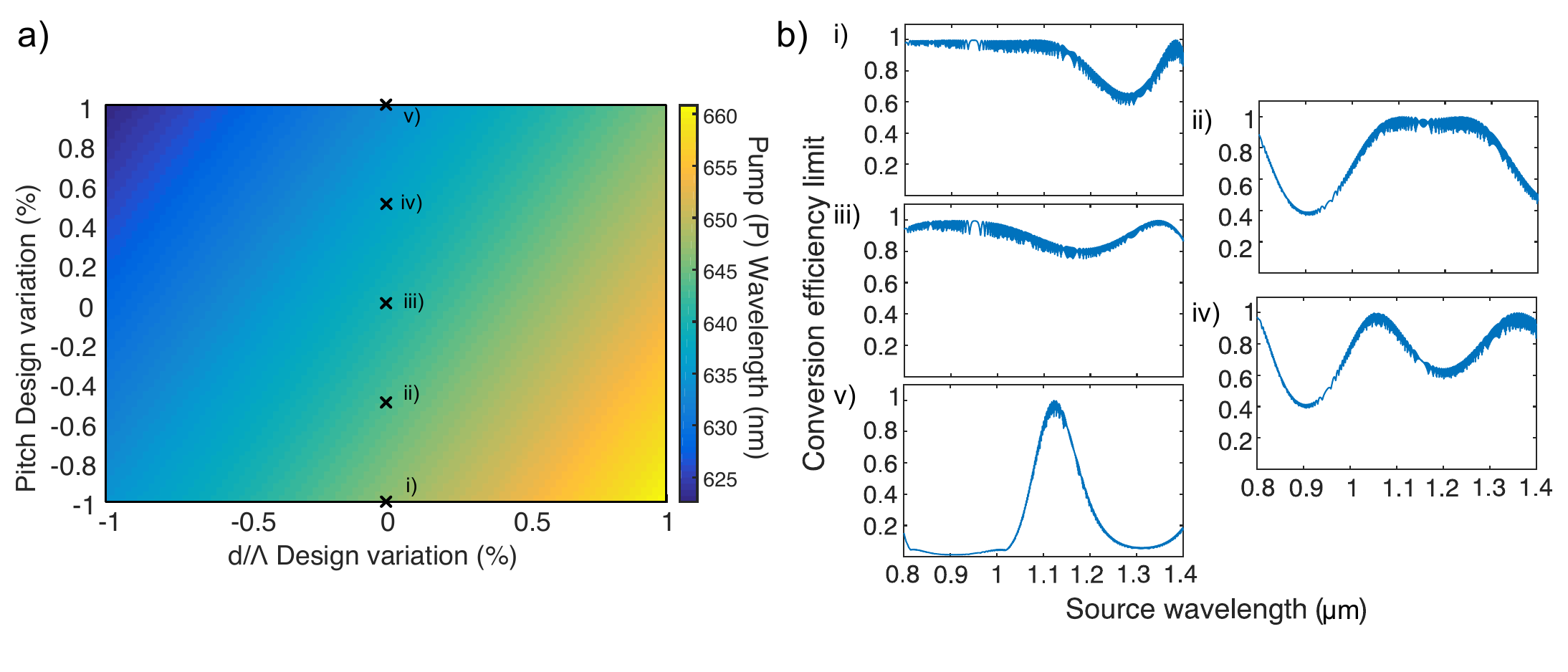}
\caption{(a) Shift in the required fixed pump wavelength, $\lambda_p$, to compensate for a $\pm$1\% change in PCF pitch (y-axis) and $d/\Lambda$ (x-axis), while maintaining the target wavelength at 1550nm.(b) Modification to conversion efficiency limit as a result of changes in pitch around the ultra-tunable PCF design discussed in Fig.\ref{fig:sym_vs_non}. Sub-panels correspond to points (i) to (v) in panel (a), from +1\% pitch in (i), through nominal PCF parameters in (iii), to -1\% pitch in (v) with $\lambda_\text{p}$ adjusted in each case to the value indicated in (a).}
\label{fig:designvariation}
\end{figure}

In order to access the ultra-tunable capabilities of this BS-FWM conversion scheme, it is necessary to match precisely the fixed pump wavelength $\lambda_\text{p}$ to the PCF parameters. If this is not achieved, the loci of points in Fig.\,\ref{fig:sym_vs_non}(c) will not intersect and efficient BS-FWM will not be achieved. In Fig.\,\ref{fig:designvariation} we analyse the effects the inevitable fabrication discrepancies in the PCF parameters will have on phase matching in a nominal ultra-tunable BS-FWM PCF design. This is parameterised in terms of the change in the value of $\lambda_\text{p}$ required to compensate for the shift in phase matching and achieve frequency conversion to the target wavelength. Here we consider discrepancies in the target parameters of pitch and $d/\Lambda$ in a typical PCF fabrication cycle; for example up to $\pm$1\% was measured in Ref.~\cite{Francis-Jones2016Characterisation-of-longitudinal-variation}, though we anticipate the designs presented here would achieve improved tolerances as the larger diameter of the holes yields better control in fabrication \cite{Wadsworth2005Hole-inflation-and-tapering}. The results in Fig.\,\ref{fig:designvariation} emphasise that PCF designs can remain within the symmetric regions indicated in Fig.\,\ref{fig:symmetric} subject to the anticipated small discrepancies in fabrication, and that BS-FWM operation can be maintained across a wide range of source wavelengths due to the capability of the fixed pump wavelength $\lambda_\text{p}$ to be tuned to achieve phase matching to the desired target wavelength.

\section{Conclusions}

We have presented a set of PCF designs within which only one fibre is required for BS-FWM frequency conversion of photons from an ultra-broad range of source wavelengths to a narrow target wavelength band. This is achieved through designing a group velocity profile that is symmetric in frequency and allows highly adaptable frequency conversion devices to be built. Furthermore, we have discussed how the wavelength of one pump laser can be tuned to correct for realistic discrepancies in fabrication while maintaining ultra-tunable operation. We expect that the resulting device, which can act as a universal quantum frequency interface, will be suitable for widespread application in unifying wavelength dispersion in quantum emitters and for deployment in distributed quantum networks.

\section{Acknowledgements}

We thank Alex Davis for comments on the manuscript. This work was supported by the UK Hub in Quantum Computing and Simulation, part of the UK National Quantum Technologies Programme with funding from UKRI EPSRC grant EP/T001062/1.

\bibliographystyle{unsrt}
\bibliography{sample}

\end{document}